\documentclass{PoS}
\usepackage{psfrag}

\title{Confining string beyond the free approximation: the case of random percolation }

\ShortTitle{Confining string beyond the free approximation: the case of random percolation }

\author{\speaker{Pietro Giudice}, Ferdinando Gliozzi, Stefano Lottini\\
        Dipartimento di Fisica Teorica, Universit\`a di Torino and INFN, sezione di Torino, Italy\\
        E-mail: \email{giudice@to.infn.it},
        \email{gliozzi@to.infn.it},
        \email{lottini@to.infn.it}
       }

\abstract
{
The random percolation model can be viewed as the dual of a 
well defined confining gauge theory;
since this theory, having no Monte Carlo dynamics at all,
is simple to simulate, it is possible to study the properties 
of the flux tube with very high precision; we show 
it can be described by the effective 
string picture. Our results are lattice regularisation 
independent, therefore they are well defined 
also in the continuum limit, and, for the first time 
in a gauge theory, it has been possible to determine the 
next-to-leading quantum corrections throughout the 
computation of the $T^6$ coefficient of the Taylor 
expansion of $\sigma(T)$. Furthermore, this coefficient 
results to be related to the universal ratio 
$T_c/\sqrt{\sigma_0}$.
}

\FullConference{The XXVI International Symposium on Lattice Field Theory \\
		 July 14 - 19, 2008\\
		 Williamsburg, Virginia, USA}

\newcommand{\eq}{\begin{equation}}
\newcommand{\en}{\end{equation}}
\newcommand{\bra}{\langle}
\newcommand{\ket}{\rangle}
\newcommand{\avg}[1]{\langle \hspace{0.2em} #1 \hspace{0.2em} \rangle}
\newcommand{\bea}{\begin{eqnarray}}
\newcommand{\ea}{\end{eqnarray}}

\begin{document}

\section{Introduction: the effective string theory}
The assumption behind the effective string theory is that the color flux
connecting a pair of quark is squeezed  inside a thin flux tube; 
as a consequence, the confining potential rises linearly. According to
this picture the flux tube should behave as a free vibrating string.

Unfortunately, the action of this effective theory is unknown; 
the simplest assumpion is that it is described by the Nambu-Goto action,
which is proportional to the string worldsheet area.

We try to summarize the outcome of many studies on this argument discussing
some properties of the first terms of the low temperature expansion of 
the string tension (the first term $\sigma_0$ is the zero-temperature 
string tension):
\eq
\sigma(T)=\sigma_0-(d-2)\frac\pi6 T^2+\sum_{n\ge3}c_nT^n \; .
\label{sexpa}
\en 
The second term, the analogue of the L\"uscher term at finite 
temperature which was calculated in Ref.~\cite{Pisarski:1982cn}, 
does not depend on the gauge group 
and is expected to be independent of the interaction terms of the 
effective theory. Thanks to a certain open-closed string 
duality it was shown that for any 
dimensionality $c_3=0$ and, in three dimensions, $c_4$ is 
universal~\cite{Luscher:2004ib}; hence, it coincides with the value 
calculated in the NG model~\cite{Arvis:1983fp,Dietz:1982uc}:
\eq
c_4=-(d-2)\frac{\pi^2}{72\,\sigma_0 } \; .
\label{c4}
\en
Using a different approach to the effective string 
theory, Ref.~\cite{Polchinski:1991ax}, the above results were confirmed
for all values of $d$.

In this paper we will evaluate the coefficients $c_n$ up to $n=6$, in a 
simple, but not trivial, model: the gauge theory dual to the $3d$
percolation model. All of the date agree with the universal values of $c_2$
and $c_4$ and lead to $c_5=0$ and 
$c_6=\pi^3/\left(C \sigma_0^2 \right)$, where $C \simeq 300$.

We decided to focus our attention to the behaviour of the Polyakov-Polyakov 
correlation function $\bra P(0)P^*(r)\ket$ at finite temperature 
$T=\frac{1}{a \ell}$ in $(2+1)$-dimensions; $r$ is the distance between the
Polyakov loops, $\ell$ is the time extent of the lattice and $a$ is the 
lattice spacing.

The functional form of the correlator has been calculated at the 
next-to-leading order (NLO) in Ref.~\cite{Dietz:1982uc}:
\eq
\label{pp}
\bra P(0)P^*(r)\ket_{NLO} = 
\frac{e^{-\mu\ell-\tilde\sigma r\ell}}{\eta(\tau)^{d-2}}\left(1-
\frac{(d-2)\pi^2\ell
[2E_4(\tau)-E_2^2(\tau)]}{1152\tilde\sigma r^3}+O\left(\frac1{r^5}\right)
\right) \; ,
\en
where $\eta$ is the Dedekind function, $E_4$ and $E_2$ the two Eisenstein 
functions and $\tau=i\ell/(2r)$.

Using Eq.~(\ref{pp}) one can find, for asymptotically large $r$:
\eq
\sigma(T)=\tilde\sigma-\frac\pi6 T^2-\frac{\pi^2}{72\tilde\sigma}T^4=
\sigma_0-\frac\pi6 T^2-\frac{\pi^2}{72\sigma_0}T^4+O(T^5)~.
\label{sigmaT}
\en
Note here the difference between $\tilde\sigma$ and $\sigma_0$: 
$\tilde\sigma=\sigma_0+O(T^5)$.

\section{The gauge theory dual to the percolation model}
In this paper we study a particular gauge theory, first introduced 
in Ref.~\cite{Gliozzi:2005ny}, that is dual to the 
random percolation model. A more complete account will be presented 
in Ref.~\cite{newpaper}.

It is known in three dimensions it is possible
to study a well defined $S_Q$-gauge\footnote{$S_Q$ is the symmetric group.}
 theory dual to the $Q$-state Potts model
through the Kramers-Wannier duality~\cite{Kramers:1941kn}. 
Thanks to the fact one can map some gauge invariant observables, 
such as Wilson loops and Polyakov correlators, into the corresponding 
quantities of the spin model, it is numerically convenient to inspect the 
properties of the dual theory instead of those of the gauge model.
The ingredient which is then fundamental in our approach
is the Fortuin-Kasteleyn reformulation~\cite{Fortuin:1971dw} of the 
$Q$-state Potts model, by which it is possible to determine gauge observables 
in a very efficient way.
This approach, that for $Q>1$ is only a powerful numerical method,
can be applied to the random percolation model whose 
gauge formulation is not known: it is the gauge theory in the $Q \to 1$ limit.

The key ingredient is the method used to calculate the Wilson loops in this 
setup:
we define a procedure to determine its value studying some
topological properties of the dual model.

The connected components of the graph, formed by active links, are known 
as clusters; $W_\gamma$ is the value of the Wilson loop associated with 
a loop with contour $\gamma$. We set  $W_\gamma=1$ if there is no cluster
topologically linked to the contour $\gamma$, otherwise we set  $W_\gamma=0$.
The same linking properties are used to determine the Polyakov-Polyakov loop
correlators $\bra P(0)P^*(r)\ket$: at finite temperature the 
contour $\gamma$ is a $r \times \ell$ rectangle, with two sides identified.

Another interesting study of this model, related to the monopole mass, can 
be found in~\cite{lottini}.

\section{Simulations}

The idea behind this work is not only to verify whether one can observe
the presence of shape effect due to rough fluctuations of the string,
in agreement with the universality predictions of the effective string 
picture (as a matter of fact we have discussed this point 
in Ref.~\cite{Giudice:2007rb}); we also would verify that our results are not 
regularisation dependent. In other words, we would discuss if our results
describe a ``real'' phenomenon and not a lattice/model artifact.
\TABULAR{|c|c|c|}{
\hline
Lattice&$p$&$\ell_c=1/aT_c$ \\
\hline 
SC bond& 0.272380&6\\
SC bond& 0.268459&7\\
SC bond& 0.265615&8\\
SC site& 0.3459514&7\\
BCC bond & 0.21113018&3\\
\hline
}
{The five systems simulated.\label{Table:1}}
We therefore study five different systems (see Table~\ref{Table:1})
characterized by different occupancy probability $p$, different kind of 
percolation (site or bond) and different geometry of the lattice 
(simple cubic lattice (SC) and body-centered cubic lattice (BCC)).

We worked on a lattice of size $L^2 \times \ell$, where $\ell$, the inverse
of the temperature, was chosen such that $0.3 T_c \lesssim  T \lesssim0.8 T_c$.
The value of the spatial size was $L=128$ which was in most cases sufficient 
to account for the infinite volume limit. Just in the case $\ell_c=8$,
simulated at $\ell=10$ and $\ell=11$, 
we found a sizable dependence on the lattice size $L$; in this case we 
performed further simulations on larger lattices in order to extract the 
correct value of $\tilde\sigma$ using the scaling relation ($\nu=4/3$ is the 
termal exponent of $2d$ percolation model):
\eq
\tilde\sigma_{1/L}=\tilde\sigma-c\,L^{-1/\nu} \; .
\en

For each system, we measured $\avg{P(0)P^*(r)}$ by varying the distance 
between the two Polyakov lines from $r=8$ to $r=50$; to reach an 
acceptable statistics, we collected data from $10^5$ configurations for 
each value of $p$ and $\ell$.

The algorithm used, described in detail in Ref.~\cite{Gliozzi:2005ny}, is
basically aimed at determining the linking properties of clusters with
the Polyakov-Polyakov contour.  

\section{Numerical results}

\begin{figure}[ht]
\center
\psfrag{sigma}{\large $\tilde\sigma$}
\includegraphics[width=10cm]{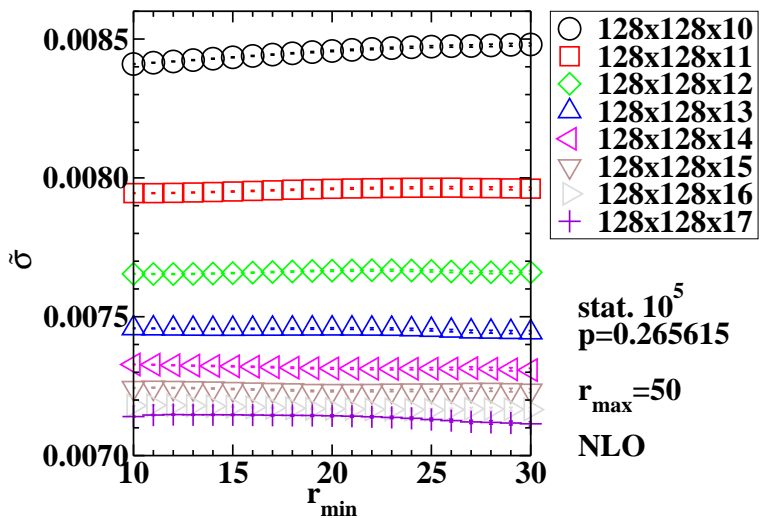}
\caption{The fitted value of the string tension $\tilde\sigma$ as a function 
of the minimal distance $r_{min}$ of the set of Polyakov-Polyakov correlators 
considered in the fit; case of bond percolation with $\ell_c=8$.}
\label{plot_plateau}
\end{figure}

\begin{figure}[ht]
\center
\psfrag{sigma}{\large $\tilde\sigma$}
\psfrag{elmenus6}{\large $\ell^{-6}$}
\includegraphics[width=10cm]{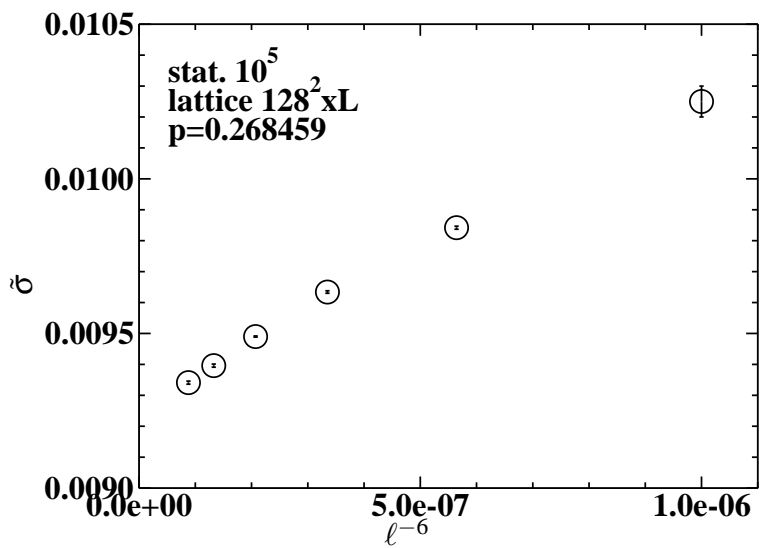}
\caption{Plot of the fitting parameter $\tilde\sigma$ as a function of $T^6$ in 
numerical experiments with bond percolation with $\ell_c=7$.}
\label{plot_sigma_Lm6}
\end{figure}

Our numerical results are compared with the expected behaviour of the 
Polyakov-Polyakov correlation function given in Eq.~(\ref{pp}).
Being an expression valid in the infrared limit 
we use a \emph{sliding window} analysis to determine the correct values of the
fitted parameter $\tilde\sigma$: we fitted the data in the range
$r_{min}\le r\le r_{max}$  by progressively discarding the short distance 
data, varying $r_{min}$ but fixing the value
of $r_{max}=50 a$ (see Fig.~\ref{plot_plateau}).
In all five sistems considered a large plateau appears for all values of
$\ell$ not too close to $\ell_c$,  showing the stability of the fits and so
the suitability of the string picture to describe our data.
It is important to note that, as  Fig.~\ref{plot_plateau} shows, there are
different values of the string tension for different values of $\ell$, i.e.
of $T$. In other words, the value of $\tilde\sigma$ is not yet the string 
tension at zero-temperature $\sigma_0$; the formula  Eq.~(\ref{pp})  is not
the exact formula because it only takes into account the temperature 
dependence up to the order $T^4$ (see Eq.~(\ref{sigmaT})).
We studied the dependence of $\tilde\sigma$ on $\ell$ and we verified,
in all cases, that for $a T=1/ \ell$ low enough the correction is 
proportional to $T^6$ (see Fig.~\ref{plot_sigma_Lm6}).
Therefore, we used the value of $\tilde\sigma$ to determine the value 
of $\sigma(T)$ by Eq.~(\ref{sigmaT}), i.e. we reconstructed the correct 
dependence of the string tension on the temperature; then we used these
data to perform a new fit to determine the first model-dependent term by 
means of the Ansatz:
\eq
\sigma(T)=\sigma_0-\frac\pi6 T^2-\frac{\pi^2}{72\sigma}T^4+
\frac{\pi^3}{C\sigma_0^2}T^6+O(T^8)~.
\label{fit}
\en
Thereby, we can identify stable values both for the zero-temperature 
string tension $\sigma_0$ and the coefficient $C$, see Table~\ref{Table:2}.
\TABULAR{|c|c|c|c|c|}{
\hline
Lattice&$\ell_c=1/aT_c$&$C$&$a^2\sigma_0$&
$T_c/\sqrt{\sigma_0}$\\
\hline 
SC bond&6&291(7)&0.012612(6)&1.4841(4)\\
SC bond&7&281(5)&0.009234(5)&1.4866(5)\\
SC bond&8&297(5)&0.007059(5)&1.4878(5)\\
SC site&7&307(9)&0.009399(8)&1.4735(6)\\
BCC bond&3&295(14)&0.0474(4)&1.531(7)\\
\hline
}
{The parameters $C$ and $a^2\sigma_0$ in the fit (\ref{fit})  
for the numerical experiments listed in 
Table \ref{Table:1}. The last column is the universal ratio
$T_c/\sqrt{\sigma_0}$ as obtained by combining the second and the fourth
columns. \label{Table:2}}

Note that the five values of $C$ coincide up to the statistical errors.
The value of $T_c/\sqrt{\sigma_0}$, obtained by combining the precise 
determination
of $a^2 \sigma_0$ with the deconfined temperaure $T_c$, is an important 
universal quantity which characterizes the particular gauge theory; 
the small variations appearing in Table~\ref{Table:2} are presumably due to the
corrections-to-scaling that we have neglected. Nonetheless, we can 
assert the value closer to the continuum limit is that obtained in the 
simulation
with bond percolation and $\ell_c=8$ where statistical and systematic error
were better under control; therefore we will use, in the following, 
$T_c/\sqrt{\sigma_0}=1.4878(5)$.

\begin{figure}[ht]
\center
\psfrag{sigmadivTsq}{\large $\sigma(T)/T^2_c$}
\psfrag{reducedT}{\large $(T-T_c)/T_c$}
\includegraphics[width=10cm]{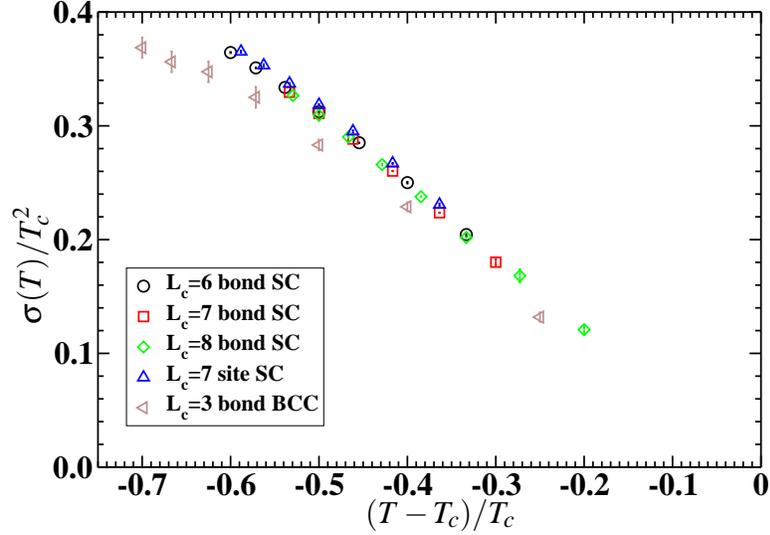}
\caption{Plot of the scaling variable $\sigma(T)/T^2_c$ versus the 
reduced temperature.}
\label{plot_adi_org}
\end{figure}

\begin{figure}[ht]
\center
\psfrag{sigmadivTsq}{\large $\sigma(T)/T^2_c$}
\psfrag{reducedT}{\large $(T-T_c)/T_c$}
\psfrag{label1}{\large $T_c/\sqrt{\sigma_0}=1.4878$}
\includegraphics[width=10cm]{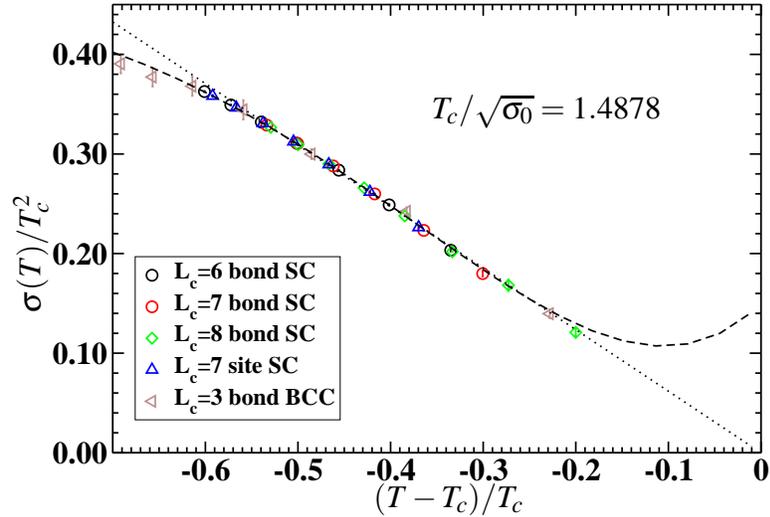}
\caption{Plot of the scaling variable $\sigma(T)/T^2_c$ versus the 
reduced temperature when we impose the value of 
$T_c/\sqrt{\sigma_0}=1.4878$. Dashed line is the plot of Eq.~(4.2),
dotted line is that of  Eq.~(4.3).}
\label{plot_adi_mod}
\end{figure}

If we plot the adimensional ratio $\sigma(T)/T_c^2$ versus the reduced 
temperature $t=(T-T_c)/T_c$ it turns out that all data lie \emph{almost} 
on a unique curve, see Figure~\ref{plot_adi_org}; this non-universal
behaviour is related to the fact the five different systems are characterized 
by different ``universal'' value of the quantity $T_c/\sqrt{\sigma_0}$
and the adimensional variables used are very sensitive to it. 
As a matter of fact, if we impose the value of $\sigma(T)/T_c^2$ is the same 
for all systems, i.e. we determine for each system a new $T_c$ value by which
$T_c/\sqrt{\sigma_0}=1.4878(5)$, all data fall on a unique universal curve
as   Figure~\ref{plot_adi_mod} shows. This is the most important result of
this work because it shows our results are independent of the
regularisation used, therefore we are studing a ``real'' gauge theory well 
defined in the continuum limit.

It is interesting to note that the values of $C$ and of $T_c/\sqrt{\sigma_0}$ 
can be determined with only two pieces of
information: (1) the data are all in the scaling region and (2) they
show a linear behaviour in the range $-0.55<t<-0.225$. 
This means we can impose the two following equations to coincide in
the above range ($S=\frac{\sigma_0}{T_c^2}$ and $x=\frac{T}{T_c}$):
\bea
\frac{\sigma(L)}{T_c^2} &=& S - \frac{\pi}{6} x^2  - \frac{\pi^2}{72 S} x^4
+ \frac{\pi^3}{C S^2} x^6 \, , \label{eqA}\\
\frac{\sigma(L)}{T_c^2} &=&A(x-1). \label{eqB}
\ea
Immediately, without using numerical data, it is possible to 
determine $C\simeq 290$ and $T_c/\sqrt{\sigma_0} \simeq 1.4884$; 
these two values are remarkably close to those obtained using the 
numerical data, see Table~\ref{Table:2}.
In Figure~\ref{plot_adi_mod} we plot Eq.~(\ref{eqA}) (dashed line) 
and Eq.~(\ref{eqB}) (dotted line) using those values; the numerical
data lie on the two curves in the scaling region.
This is an important observation because it means the two quantity, 
$C$ and $T_c/\sqrt{\sigma_0}$ are constrained to each other in that region.

\section{Conclusions}

In this paper we have studied, by numerical simulation, the gauge theory 
dual to the percolation model; we can conclude it is possible
to describe the long distance dynamics of this theory by means of an
effective string picture. 
Our numerical experiment demonstrate that the quantities which characterize 
the effective string theory do not depend on the specific regularisation used.
Moreover, we determined with high precision 
the value of $\sigma(T)/T_c^2$ and, for the first time
in a gauge theory, we have determined the value of the $T^6$ coefficient $C$
of the string tension $\sigma(T)$. Furthermore, it was possible to show
that the universal ratio $T_c/\sqrt{\sigma_0}$ and the coefficient $C$ are 
bound together in the scaling region.

\end{document}